\documentclass[12pt]{article}
\def\one{
\setlength{\unitlength}{0.45cm}
\begin{picture}(0.55,0.5)
\put(0,0){\line(1,0){0.4}}
\put(0,.4){\line(1,0){0.4}}
\multiput(0,0)(.4,0){2}{\line(0,1){.4}}
\end{picture}}
\def\twohor{
\setlength{\unitlength}{0.45cm}
\begin{picture}(1.1,0.5)
\put(0,0){\line(1,0){0.8}}
\put(0,.4){\line(1,0){0.8}}
\multiput(0,0)(.4,0){3}{\line(0,1){.4}}
\end{picture}}
\def\twover{
\setlength{\unitlength}{0.45cm}
\begin{picture}(0.55,0.5)
\put(0,0){\line(1,0){0.4}}
\put(0,.4){\line(1,0){0.4}}
\put(0,-.4){\line(1,0){0.4}}
\multiput(0,0)(.4,0){2}{\line(0,0){.4}}
\multiput(0,0)(.4,0){2}{\line(0,-1){.4}}
\end{picture}}
\def\threehor{
\setlength{\unitlength}{0.45cm}
\begin{picture}(1.5,0.5)
\put(0,0){\line(1,0){1.2}}
\put(0,.4){\line(1,0){1.2}}
\multiput(0,0)(.4,0){4}{\line(0,1){.4}}
\end{picture}}
\def\threever{
\setlength{\unitlength}{0.45cm}
\begin{picture}(0.6,0.5)
\put(0,0){\line(1,0){0.4}}
\put(0,.4){\line(1,0){0.4}}
\put(0,-.4){\line(1,0){0.4}}
\put(0,-.8){\line(1,0){0.4}}
\multiput(0,0)(.4,0){2}{\line(0,0){.4}}
\multiput(0,0)(.4,0){2}{\line(0,-1){.4}}
\multiput(0,0)(.4,0){2}{\line(0,-2){.8}}
\end{picture}}
\def\mixed{
\setlength{\unitlength}{0.45cm}
\begin{picture}(1,0.5)
\put(0,0){\line(1,0){0.8}}
\put(0,.4){\line(1,0){0.8}}
\put(0,-.4){\line(1,0){0.4}}
\multiput(0,0)(.4,0){3}{\line(0,1){.4}}
\multiput(0,0)(.4,0){2}{\line(0,-1){.4}}
\end{picture}}
\def\fourhor{
\setlength{\unitlength}{0.45cm}
\begin{picture}(2.2,0.5)
\put(0,0){\line(1,0){1.6}}
\put(0,.4){\line(1,0){1.6}}
\multiput(0,0)(.4,0){5}{\line(0,1){.4}}
\end{picture}}

\usepackage{epsf}
\usepackage{cite}
\def\one{
\setlength{\unitlength}{0.45cm}
\begin{picture}(0.55,0.5)
\put(0,0){\line(1,0){0.4}}
\put(0,.4){\line(1,0){0.4}}
\multiput(0,0)(.4,0){2}{\line(0,1){.4}}
\end{picture}}
\def\twohor{
\setlength{\unitlength}{0.45cm}
\begin{picture}(1.1,0.5)
\put(0,0){\line(1,0){0.8}}
\put(0,.4){\line(1,0){0.8}}
\multiput(0,0)(.4,0){3}{\line(0,1){.4}}
\end{picture}}
\def\twover{
\setlength{\unitlength}{0.45cm}
\begin{picture}(0.55,0.5)
\put(0,0){\line(1,0){0.4}}
\put(0,.4){\line(1,0){0.4}}
\put(0,-.4){\line(1,0){0.4}}
\multiput(0,0)(.4,0){2}{\line(0,0){.4}}
\multiput(0,0)(.4,0){2}{\line(0,-1){.4}}
\end{picture}}
\def\threehor{
\setlength{\unitlength}{0.45cm}
\begin{picture}(1.5,0.5)
\put(0,0){\line(1,0){1.2}}
\put(0,.4){\line(1,0){1.2}}
\multiput(0,0)(.4,0){4}{\line(0,1){.4}}
\end{picture}}
\def\threever{
\setlength{\unitlength}{0.45cm}
\begin{picture}(0.6,0.5)
\put(0,0){\line(1,0){0.4}}
\put(0,.4){\line(1,0){0.4}}
\put(0,-.4){\line(1,0){0.4}}
\put(0,-.8){\line(1,0){0.4}}
\multiput(0,0)(.4,0){2}{\line(0,0){.4}}
\multiput(0,0)(.4,0){2}{\line(0,-1){.4}}
\multiput(0,0)(.4,0){2}{\line(0,-2){.8}}
\end{picture}}
\def\mixed{
\setlength{\unitlength}{0.45cm}
\begin{picture}(1,0.5)
\put(0,0){\line(1,0){0.8}}
\put(0,.4){\line(1,0){0.8}}
\put(0,-.4){\line(1,0){0.4}}
\multiput(0,0)(.4,0){3}{\line(0,1){.4}}
\multiput(0,0)(.4,0){2}{\line(0,-1){.4}}
\end{picture}}

\catcode`@=11 \@addtoreset{equation}{section} \catcode`@=12

\begin{document}
\begin{titlepage}
\vfill
\begin{center}
{\Large \bf $U(N)$ Framed Links, Three-Manifold Invariants, and 
Topological Strings }\\[1cm] 
Pravina Borhade \footnote{E-mail: pravina@phy.iitb.ac.in}, 
P. Ramadevi\footnote{Email: ramadevi@phy.iitb.ac.in}\\
{\em Department of Physics, \\Indian Institute of Technology Bombay,\\
Mumbai 400 076, India\\[10pt]}
Tapobrata Sarkar\footnote{Email: tapo@ictp.trieste.it}\\
{\em the Abdus Salam International Centre for Theoretical Physics, 
\\ Strada Costiera, 11 -- 34014, Trieste, Italy}\\ 
\end{center}
%\vspace{2cm}
\vfill
\begin{abstract}
Three-manifolds can be obtained through surgery of framed links in $S^3$.
We study the meaning of surgery procedures
in the context of topological strings. 
We obtain $U(N)$ three-manifold invariants from $U(N)$ 
framed link invariants in Chern-Simons theory on $S^3$. 
These three-manifold invariants are proportional to the 
Chern-Simons partition function on the respective 
three-manifolds. Using the topological string duality conjecture, we 
show that the large $N$ expansion of $U(N)$ Chern-Simons free energies 
on three-manifolds, obtained from some class of framed links, have a closed
string expansion. These expansions resemble the closed string $A$-model 
partition functions on Calabi-Yau manifolds with one
Kahler parameter. We also determine Gopakumar-Vafa integer coefficients 
and Gromov-Witten rational coefficients corresponding to Chern-Simons
free energies on some three-manifolds.
\end{abstract}
\vfill
\end{titlepage}

\section{Introduction}
After the second superstring revolution, several useful
relations have been discovered unifying various ideas of physics and
mathematics. One such surprising discovery in the recent past has
been the new connections between Chern Simons gauge theory and the physics
of closed topological string theory in certain backgrounds.

The initial steps in this direction was taken by Gopakumar and Vafa in
\cite{gv1},\cite{gv2},\cite{gv3}. The conjecture put forward by these
authors relate large $N$ Chern-Simons gauge theory on $S^3$,
which is equivalent to A-twisted open topological string theory on
$T^*S^3$ \cite{wittencs}, to the A-type closed topological string theory 
on the resolved conifold. This conjecture was then tested at the 
level of the observables of the Chern Simons theory, namely the 
knot invariants. In \cite{ov}, Ooguri and Vafa
formulated the conjecture in terms of invariants for the unknot (a circle
in $S^3$), and further checks were carried out for more nontrivial knots in
\cite{lm},\cite{taps},\cite{laba},\cite{mari}.

The evaluation of knot invariants in \cite{ov} actually led to very strong
integrality predictions for the (instanton generated) A-model disc amplitudes,
which were then verified from the more tractable mirror B-model side by
several authors \cite{av},\cite{akv}, \cite{gjs},\cite{lermyw}.

Purely from gauge theory considerations, following the idea of `t Hooft
\cite{thoo}, it looks to be a challenging problems to prove that the
Feynmann perturbative expansion of any $U(N)$ gauge theory in the large $N$
limit is equivalent to a closed string theory. It is believed that
the Gopakumar-Vafa duality conjecture can provide insight in determining
the `t Hooft expansion of $U(N)$ Chern-Simons free energy on any
three-manifold $M$.

As we have already mentioned, the Gopakumar-Vafa duality conjecture states
that $U(N)$ Chern-Simons theory on $S^3$, which describes the
topological $A$-model of $N$ D-branes on $X=T^*S^3$, is dual
to topological closed string theory on
$X^t= {\cal O}(-1) \oplus {\cal O}(-1) \rightarrow {\cal P}^1$.
Having verified the conjecture at the level of Chern-Simons partition
function on $S^3$ and Wilson loop observables (the knot invariants), we
need to understand the meaning of surgery of framed links in $S^3$
within the context of topological strings, and this is one of the issues
we set out to address in this paper.

From the fundamental theorem of Lickorish and Wallace \cite{lickorish}, 
it is well known that any three-manifold $M$ can be obtained
by surgery on a framed link in $S^3$. Further, two framed links
related by a set of moves called Kirby moves determine
the same manifold. In Chern-Simons theory, an algebraic expression
has been derived \cite{kaul}, in terms of framed link invariants, 
which are unchanged under Kirby moves of the framed links.  
Hence the algebraic expression represents three-manifold invariants which 
are proportional to the Chern-Simons partition function ($Z[M]$) on the 
three-manifold $M$. Incorporating the results of the 
topological string duality conjecture, we determine large 
$N$ expansion for ${\rm ln} Z[M]$ for many manifolds. 
Surprisingly, the expansion looks like an $A$-model closed 
string partition function on a Calabi-Yau space with one Kahler parameter.

We point out an important subtelty here. As is well known, the
classical solutions of the Chern-Simons action on a general 
theree-manifold $M$ are the flat connections on $M$. In the weak 
coupling limit, the Chern-Simons partition function gets contributions 
from a perturbative expansion around all such stationary points. 
Noting that the space of 
flat connections may be either a collection of a set of stationary points 
or a set of connected pieces, this partition function can be appropriately 
written as a sum or an integral over the space of flat connections.
The large $N$ expansion of 't Hooft is expected to relate the $1/N$  
expansion of the Chern-Simons theory around a given flat connection to
an A-type closed topological string theory. In \cite{akmv}, this has 
been shown from a matrix model approach, for the Lens space ${\cal L}(p,1)$. 
In this paper, however, we show that the full Chern-Simons partition
function $({\rm ln} Z[M])$ has a closed string interpretation for 
a class of three manifolds. 

Indeed, proposing new duality conjectures between Chern-Simons 
theory on general three-manifolds $M$ and the corresponding 
dual closed string theories will involve the extraction of the
partition function around individual flat connections, in lines
with \cite{akmv}. This will extend the original conjecture by Gopakumar
and Vafa, for general manifolds $M$. We believe that our results on the 
invariants, involving the full partition function, would be 
useful in proposing and understanding fully the nature of such dualities. 
We will elaborate on this point further in the concluding section.  

The organisation of the paper is as follows. In
section 2, we briefly recapitulate the framed link invariants
in $U(N)$ Chern-Simons theory, and present the
$U(N)$ three-manifold invariants obtained from framed
link invariants in $S^3$. In section 3, we show the relation between the 
three-manifold invariants and the observables in topological string theory. 
Further, we obtain closed string invariants for the Chern-Simons free energies.
Section 4 contains some explicit results on the Gopakumar Vafa coefficients
corresponding to the large $N$ expansion of the Chern-Simons free energy 
on some manifolds. Section 5 ends with some discussions and
scope for future research. In an appendix, we present some results
on the integer invariants for the unknot with arbitrary framing, which are
useful for the computation of the Gopakumar Vafa coefficients.

\section{$U(N)$ Chern-Simons Gauge theory}

Chern-Simons gauge theory on a three-manifold $M$ 
based on the gauge group $U(N)$ is a factored
Chern-Simons theory of two gauge groups, $SU(N)$ and $U(1)$. 
That is, the action is simply a sum of two Chern-Simons actions,
one for gauge group $SU(N)$ and the other for $U(1)$, each with an independent
coupling constant ($k,k_1$) 
\begin{equation}
S = {k \over 4 \pi} \int_M Tr\left (A \wedge dA + {2 \over 3} A \wedge
A \wedge A \right) + {k_1 \over 4 \pi} \int_M Tr\left (B \wedge dB\right) 
\end{equation}
where $A$ is a gauge connection for gauge group $SU(N)$ and
$B$ is the connection for $U(1)$. Clearly, the $U(N)$ partition function 
$Z_{\{U(N)\}}[M] \equiv Z[M]$ is just the product of two partition functions 
[$Z_{\{SU(N) \}}[M]$, $Z_{\{U(1)\}}[M]$] 
\begin{equation}
Z[M] = \int[{\cal D}B][{\cal D}A] ~e^{iS}~.
\end{equation}

%The classical solutions of the Chern-Simons action are flat connections 
%on $M$. In the weak coupling limit (large $k,k_1$), such 
%an integral is given by a sum of contributions from the stationary points 
%\begin{equation}
%Z_{\{U(N)\}}[M] = Z_{\{U(N)\}}^{(tr)}[M] + \sum_{c \neq (tr)} 
%Z_{\{U(N)\}}^{(c)}[M]
%\end{equation}
%where $Z_{\{U(N)\}}^{(tr)}[M]$ corresponds to the perturbative expansion
%of the path-integral around the trivial connection and  
%$Z_{\{U(N)\}}^{(c)}[M]$ will involve a perturbative 
%expansion around non-trivial connections. It is believed that
%the Feynmann perturbative expansion around any stationary point
%should have a closed string theory interpretation\cite {thoo}
%which is the central theme of the subject matter addressed in 
%later sections.

We shall now briefly present the Wilson loop observables in the theory.
The $U(N)$ Wilson loop operators for a $r$-component link $L$ made up 
of component knots ${\cal K}_i$'s are simply factored 
Wilson operators of the $U(1)$ and $SU(N)$ theories 
\begin{equation}
W_{\{(R_i,n_i)\}}[L]~=~ \prod_{i=1}^rTr_{R_i} U^{(A)} [{\cal K}_i]~ 
Tr_{n_i} U^{(B)}[{\cal K}_i]~,
\end{equation}	
where $U^{(A)}[{\cal K}_i]=P\left[\exp \oint_{{\cal K}_i} A\right]$ denotes 
the holonomy of the 
$SU(N)$ gauge field $A$ around the component knot ${\cal K}_i$
of a link $L$  carrying 
representation $R_i$ and $U^{(B)}[{\cal K}_i]=
P\left[\exp \oint_{{\cal K}_i} B\right]$ 
denotes the holonomy of the 
$U(1)$ gauge field $B$ around the component knot ${\cal K}_i$ carrying 
$U(1)$ charge $n_i$.    
The expectation value of these Wilson loop operators are the
$U(N)$ link invariants which are products of $SU(N)$ and $U(1)$ invariants 
\begin{eqnarray}
V^{\{U(N)\}}_{(R_1,n_1),(R_2,n_2), \ldots (R_r,n_r)}[L,M]&=& 
\langle W_{\{(R_i,n_i)\}}[L] \rangle= {\int[{\cal D}A][{\cal D} B]e^{iS}
 W_{\{(R_i,n_i)\}}[L] \over \int[{\cal D} A][{\cal D} B]e^{iS}} \nonumber\\
~&=&V^{\{SU(N)\}}_{R_1,R_2, \ldots R_r}[L,M]~ 
V^{\{U(1)\}}_{n_1,n_2, \ldots n_r}[L,M]
 \label {linki} 
\end{eqnarray}

\subsection{$U(N)$ Framed Link Invariants in $S^3$}
The observables in $U(1)$ Chern-Simons theory on a three-sphere $S^3$ 
capture only self-linking numbers (also called framing numbers) and the 
linking numbers between the component knots of any link. Hence, 
the $U(1)$ link invariant will be 
\begin{equation}
V^{\{U(1)\}}_{n_1,n_2, \ldots n_r}[L,S^3]~= \exp \left({i \pi \over k_1}
{\sum_{i=1}^r~n_i^2 p_i} \right)
\exp \left({i \pi \over k_1}{\sum_{i \neq j} n_i n_j \ell k_{ij}} \right)~. 
\label {linkj}
\end{equation}
where $p_i$'s are the framing numbers of the component knots 
${\cal K}_i$'s and $\ell k_{ij}$ are
the linking numbers between the component knots ${\cal K}_i, {\cal K}_j$.
From eqns. (\ref {linki}),(\ref{linkj}), it is clear that the $U(N)$ link
invariants coincides with $SU(N)$ invariants if and only if
$p_i$'s and $\ell k_{ij}$ are zero. 

The evaluation of $SU(N)$ framed link invariants 
from Chern-Simons theory on $S^3$ makes use of
the two ingredients: (i) the connection between 
Chern-Simons field theory and the corresponding Wess-Zumino 
conformal field theory (ii) the fact that knots and links can be obtained
by closure or platting of braids. We refer the reader to 
\cite {swati}\cite{kaul} for detailed description of obtaining 
$SU(2)$ framed invariants and framed invariants from Chern-Simons 
theory on $S^3$ based on any arbitrary semi-simple group. 

We shall now present the polynomials
for various framed knots and links. For the unknot $U$ with an 
arbitrary framing $p$, carrying a representation $R$ of $SU(N)$, the 
polynomial is 
\begin{equation}
V^{\{SU(N)\}}_R[0^{(p)}, S^3] =q^{(p C_R)}~ V_R[U]=q^{-{p {\ell}^2 \over 2 N}} 
(q^{p \kappa_R} dim_q R~),
\end{equation}
where $q=\exp \left(2 \pi i \over k+N \right)$, 
$\ell$ refers to the total number of boxes in the Young-Tableau of the 
representation $R$ and 
\begin{equation}
\kappa_R= {1 \over 2} \left (N \ell + \ell + \sum_i (l_i^2- 2i l_i) \right)~,
\end{equation} 
with $l_i$ being the number of boxes in the $i$-th row
of the Young-Tableau of the representation $R$. One can verify that 
both the quantum dimension of a representation, $dim_q R$ and  $\kappa_R$ 
are polynomials in variables $\lambda^{\pm {1 \over 2}}$ and $q^{\pm {1 \over 2}}$
where $\lambda=q^N$. The frame dependent term involves a
variable $e^z=q^{1 \over 2N}$. Hence the unknot invariants with framing
$p \neq 0$ are no longer polynomials in variables $q^{\pm {1 \over 2}},
\lambda^{\pm {1 \over 2}}$ but also involve one more variable 
$e^z=q^{1 \over 2 N}$. 

In order to make the polynomials independent of 
the variable $e^z$, we can multiply by $U(1)$ invariant (\ref {linkj})
with a specific choice of $U(1)$ charge $n$ and coupling constant $k_1$ 
\begin{equation}
n  = {\ell \over \sqrt{N}} ~;~~ k_1 =k+N~. \label {charg}
\end{equation}
Therefore the $U(N)$ invariant for the $p$-framed unknot in $S^3$ with the
above choice of $U(1)$ representation and coupling constant is 
\begin{equation}
V^{\{U(N)\}}_{(R,{{\ell} \over \sqrt{N}})}[0^{(p)},S^3] = 
q^{p \kappa_R} dim_q R~.
\end{equation}
Now, we can write the $U(N)$ framed knot invariants for torus knots
of the type $(2, 2m+1)$ and other non-torus knots like $4_1$, $6_1$ 
in $S^3$. For example, the $U(N)$ invariant for a framed torus knot of
type $K \equiv (2,2m+1)$ with framing $[p - (2m+1)]$ will be 
\begin{equation}
V^{\{U(N)\}}_{(R, {{\ell} \over \sqrt N})}[K, S^3]= q^{p \kappa_R}
\sum_{R_s \in R \otimes R} dim_q R_s ~(-1)^{\epsilon_s} 
\left(q^{\kappa_R-\kappa_{R_s}/2}\right)^{2m+1}~,
\end{equation}
where $\epsilon_s= \pm 1$ depending upon whether the representation $R_s$
appears symmetricially or antisymmetrically with respect to the tensor
product $R \otimes R$ in the $SU(N)_k$ Wess-Zumino Witten model. 
Similarly, $U(N)$ invariants for framed torus links of the type $(2,2m)$
can also be written. For example, the $U(N)$ invariant for a 
Hopf link with linking number $-1$ and framing numbers 
$p_1$ and $p_2$ on the component knots carrying representions 
$R_1$ and $R_2$ will be 
\begin{equation}
V^{\{U(N)\}}_{(R_1, {\ell_1 \over \sqrt N}),
(R_2, {\ell_2 \over \sqrt N})}[H^*(p_1,p_2),S^3]= q^{p_1 \kappa_{R_1}}
q^{p_2 \kappa_{R_2}}
\sum_{R_s \in R_1 \otimes R_2} dim_q R_s q^{\kappa_{R_1} + \kappa_{R_2}- 
\kappa_{R_s}}~,
\end{equation}
where $\ell_1$ and $\ell_2$ refers to total number of
boxes in the Young-Tableau of the representations $R_1$ and $R_2$ 
respectively. From now on, we shall denote 
$$V^{\{U(N)\}}_{(R_1,{\ell_1 \over \sqrt N}),
(R_2, {\ell_2 \over \sqrt N}), \ldots (R_r, {\ell_r \over 
\sqrt N})} [L,S^3]\equiv V^{\{U(N)\}}_{R_1, R_2, \ldots R_r}[L,S^3]
$$
supressing the $U(1)$ charges as they are related to the 
total number of boxes in the Young-Tableau of the representations $R_i$'s 
(\ref {charg}). Using the framed $SU(N)$ invariants for 
arbitrary framed links in $S^3$ \cite{kaul},
it is straightforward to obtain the corresponding $U(N)$ link invariants.
We will now see how these $U(N)$ framed invariants in $S^3$
with such special $U(1)$ representation reflects 
on three-manifold invariants.

\subsection{$U(N)$ Three-Manifold Invariants}

The Lickorish-Wallace theorem states that any three-manifold $M$ can be
obtained by a surgery of framed knots and links in $S^3$. Two 
framed links related by Kirby moves will determine the same 
three-manifold. In other words, three-manifold invariants must
be constructed from framed link invariants in such a way that they are 
preserved under Kirby moves. The $SU(N)$ three-manifold invariant 
$F[M]$ for a manifold $M$ obtained 
by surgery of a framed link in $S^3$ will be \cite {kaul} 
\begin{equation}
F[M]= \alpha^{-\sigma[L]} \sum_{R_1, R_2 \ldots R_r} \mu_{R_1, R_2 \ldots
R_r}(\{p_i \}, \{\ell k_{ij} \}) V^{\{SU(N)\}}_{R_1,R_2, \ldots R_r}[L,S^3]~,
\end{equation}
where $\{p_i\}, \{\ell k_{ij}\}$ are 
the framing and linking numbers and $\sigma(L)$ is the 
signature of the linking matrix of framed link
$L$. It has been proven \cite {kaul} that $F[M]$ is unchanged under the 
operation of Kirby moves on framed links if we choose 
\begin{equation}
\alpha= \exp \left({i \pi c \over 4}\right)~,~~
\mu_{R_1, R_2 \ldots R_r}(\{p_i \}, \{\ell k_{ij} \})=
S_{0R_1} S_{0R_2} \ldots S_{0R_r} ~, 
\end{equation}
where 
$c={k(N^2-1) \over (k+N)}$ and
$S_{0R_i}$'s denotes the  modular transformation matrix 
elements. We see that $\mu_{R_1,R_2 \ldots R_r}$ is independent of
framing and linking numbers. We can now construct $U(N)$ three-manifold
invariants from the $U(N)$ framed link invariants in $S^3$ as follows 
\begin{equation}
\tilde F[M] = \beta^{-\sigma[L]} \sum_{\{R_i\}} 
{\tilde \mu}_{R_1, R_2 \ldots R_r} (\{p_i \},
\{\ell k_{ij} \}) V^{\{U(N)\}}_{R_1,R_2, \ldots R_r}[L,S^3]~,
\end{equation}
where $\beta$ and ${\tilde \mu}$ must be chosen such that
$\tilde F[M]$ is unchanged under Kirby moves on framed links.
Further, for obtaining three-manifolds from knots and disjoint links 
with zero framing and linking numbers, we require 
\begin{equation}
\tilde F[M]=F[M]~,~ {\rm as}~ V^{\{U(N)\}}_{R_1, \ldots R_r}[L,S^3]
= V^{\{SU(N)\}}_{R_1, \ldots R_r}[L,S^3]~.
\end{equation}
Therefore, for $\{p_i\}=0, \{\ell k_{ij}\}=0$ 
\begin{equation}
{\tilde \mu}_{R_1 R_2 \ldots R_r}(\{p_i=0\}, \{\ell k_{ij}=0 \})
=\mu_{R_1 R_2 \ldots R_r}~.
\end{equation}
The special choice of the $U(1)$ representation and 
the above limiting conditions suggests that
%\begin{eqnarray}
\begin{equation}
\beta =\alpha~,~~~~~
{\tilde \mu}_{R_1 R_2 \ldots R_r}(\{p_i\}, \{\ell k_{ij} \})
=\mu_{R_1 R_2 \ldots R_r} e^{-z \left (\sum_{i} l_i^2 p_i 
+\sum_{i \neq j} l_i l_j \ell k_{ij} \right)}~,
\end{equation}
for $\tilde F[M]$ to be preserved under Kirby moves. 
Rewriting the $U(N)$ link invariants in terms of the $SU(N)$ invariants,
it is obvious that $\tilde F[M]$ is the same as $F[M]$.
Hence the three-manifold invariants do not distinguish between
$U(N)$ and $SU(N)$ gauge groups and they are proportional to
the partition function $Z[M]$\cite{kaul}:
\begin{equation}
F[M] = {Z[M] \over Z[S^3]}~.
\end{equation}
% 
%
%In Ref. \cite {rozan}, \footnote{We would like to thank N. Habegger 
%for comments on this point.} a general formula for the
%Chern-Simons partition function on $M$ obtained from
%links invariants in $M'$ (not necessarily $S^3$) 
%has been presented. In the weak coupling limit, the link polynomials 
%in $M'$ will have contributions from various stationary points
%which will determine the perturbative 
%partition function on $M$ around those stationary points.
%The three-sphere $S^3$ has only a trivial flat connection 
%and the framed link invariants in $S^3$ 
%can be obtained from a perturbative expansion
%near the trivial connection. Comparing 
%the three-manifold invariants $F[M]$ from framed link invariants
%in $S^3$  with the Witten-Reshetikhin-Turaev invariant \cite {rozan}, 
%we deduce that
%\begin{equation}
%F[M] = {Z^{(tr)}[M] \over Z[S^3]}~,
%\end{equation}
%in the weak coupling limit, where $Z^{(tr)}[M]$ is the 
%partition function obtained from a perturbative
%expansion around the trivial connection. 
The partition function on $S^3$ is equal to
\begin{equation}
Z[S^3]=S_{00}~.~\label {spher}
\end{equation}
Let us introduce a slight modification in notation which will be useful
when we relate these three-manifold invariants with expectation
values of topological operators in topological string theory 
\begin{equation}
V^{\{U(N)\}}_{R_1,R_2 \ldots R_r}[L,S^3]= (-1)^{\sum_i \ell_i p_i} \lambda^{\sum_i
\ell_i p_i \over 2} \tilde V^{\{U(N)\}}_{R_1,R_2, \ldots R_r}[L,S^3](q,
\lambda)~. \label {usef}
\end{equation}
The $U(N)$ (or equivalently $SU(N)$) three-manifold invariants can be
rewritten as 
%\begin{eqnarray}
\begin{equation}
%\tilde F[M]&=& \alpha^{-\sigma[L]} 
%\sum_{\{R_i \}} \mu_{R_1,R_2 \ldots R_r}
%e^{z (\sum_i \ell_i^2 p_i+ \sum_{i,j} \ell_i \ell_j \ell k_{ij})}
%(-1)^{\sum_i \ell_i p_i} \lambda^{\sum_i \ell_i p_i \over 2} \tilde V^
%{(U(N))}_{R_1,R_2 \ldots R_r}[L,S^3](q,\lambda)\\
{Z[M] \over S_{00}}= {Z_0[M] \over S_{00}} + 
{\cal F}[M, z,\{p_i\}, \{\ell k_{ij}\}]
%\end{eqnarray}
\end{equation}
where $Z_0[M]/S_{00}$ is independent of $z$, $\{p_i\}$, $\{\ell k_{ij} \}$
and is given by 
\begin{equation}
{Z_0[M] \over S_{00}}=
 \alpha^{-\sigma[L]} \sum_{R_1,R_2, \ldots R_r} S_{0R_1} \ldots
S_{0 R_r} (-1)^{\sum_i \ell_i p_i}  \lambda^{\sum_i \ell_i p_i \over 2}
\tilde V^{\{U(N)\}}_{R_1,R_2, \ldots R_r}[L,S^3](q,\lambda)~ \label {pinde},
\end{equation}
and ${\cal F}[M,z, \{p_i\}, \{\ell k_{ij} \}]$ contains the remaining
$z$, $\{p_i\}$, $\{\ell k_{ij}\}$ dependent terms. For 
knots and disjoint links with zero framing and linking
numbers, it is not difficult to see that
${\cal F}[M,z, \{p_i \}, \{\ell k_{ij} \}] =0$ resulting
in $Z[M] = Z_0[M]$.
In the next section, we will show the natural appearance of
$Z_0[M]$ in the context of topological strings.

\section{Topological Strings} 

Gopakumar and Vafa have conjectured that closed topological string 
theory on a resolved conifold is dual to large $N$ Chern-Simons
gauge theory on $S^3$. The conjecture has been verified by comparing the 
large $N$ expansion of the free-energy of the Chern-Simons
theory on $S^3$ with the closed topological string amplitude near 
the resolved conifold. This duality
relates the Chern-Simons field theory variables 
$q$ and $\lambda$ with the string theory parameters 
\begin{equation}
q = e^{g_s}~~,~~ \lambda=e^t= e^{Ng_s}~,
\end{equation}
where $g_s$ is the string coupling constant and $t$ is the
Kahler parameter of the resolved conifold.
With the above identification between the variables $q, \lambda$
with $g_s,t$, the Chern-Simons variable $e^z$ is 
\begin{equation}
e^z= e^{g_s/2N} ~.
\end{equation}
The large $N$ expansion is performed by taking the limits 
\begin{equation}
g_s \rightarrow 0 ~ {\rm and}~ N \rightarrow \infty~.
\end{equation}
In this limit, the variable $z= g_s/2N$ can be set to 
zero. This suggests that the
$z$ independent part of the three-manifold invariant, namely, 
$Z_0[M]/S_{00}$ can be compared with quantities on the 
topological string side. 

Ooguri and Vafa found another piece of evidence for this duality
conjecture by showing that the Wilson loop operators in
Chern-Simons theory correspond to certain observables in 
the topological string theory. The operators in the open topological 
string theory which contains information
about links is given by \cite {ov} 
\begin{equation}
Z(\{U_{\alpha}\}, \{V_{\alpha}\})= \exp\left[ \sum_{\alpha=1}^r \sum_{d=1}
^{\infty} {1 \over d} {\rm Tr} U_{\alpha}^d~ {\rm Tr} V_{\alpha}^d \right]
\label {opr}
\end{equation}
where $U_{\alpha}$ is the holonomy of the gauge connection $A$ around
the component knot ${\cal K}_{\alpha}$ carrying the 
fundamental representation in the $U(N)$ Chern-Simons theory on
$S^3$, and $V_{\alpha}$ is the holonomy of a gauge field $\tilde A$
around the same component knot carrying the fundamental representation
in the $U(M)$ Chern-Simons theory on a Lagrangian three-cycle which 
intersects $S^3$ along the curve ${\cal K}_{\alpha}$. 

We can use some group theoretic properties to show that
the expectation value of the operator (\ref {opr}) exactly
matches $Z_0[M]/S_{00}$ provided we choose the rank of the 
two Chern-Simons gauge groups to be same ($N=M$). If we expand
the exponential in eqn. (\ref{opr}), we will get 
\begin{equation}
Z(\{U_{\alpha}\}, \{V_{\alpha} \})=  1+ \sum_{\{\vec 
k^{(\alpha)}\}} \prod_{\alpha=1}^r 
{1 \over z_{\vec k^{(\alpha)}}} {\bf \gamma}_{\vec k^{(\alpha})} (U_{\alpha})
{\bf \gamma}_{\vec k^{(\alpha)}} (V_{\alpha}) \label {compa}
\end{equation}
where 
\begin{equation}
z_{\vec k^{(\alpha)}}= \prod_j k_j^{(\alpha)}! j^{k_j^{(\alpha)}}~,
{\bf \gamma}_{\vec k^{(\alpha)}} (U_{\alpha}) = \prod_{j=1}^{\infty} 
\left({\rm Tr}U_{\alpha}^j \right)^{k_j^{(\alpha)}}~.
\end{equation}
Here $\vec k^{(\alpha)}=(k_1^{(\alpha)},k_2^{(\alpha)}, \ldots)$ 
with $\vert \vec k^{(\alpha)} \vert = \sum_j k_j^{(\alpha)}$ and the
sum is over all the vectors $\vec k^{(\alpha)}$ such that
$\sum_{\alpha=1}^r \vert \vec k^{(\alpha)}\vert >0$. 
Using the group theoretic properties 
\begin{eqnarray}
{\bf \gamma}_{k_1}(U_1) \ldots {\bf \gamma}_{k_r}(U_r)=\sum_{R_1, \ldots R_r} 
\prod_{\alpha=1}^r \chi_{R_{\alpha}}(C(\vec k^{(\alpha)})) 
{\rm Tr}_{R_1}(U_1) \ldots {\rm Tr}_{R_r}(U_r) ~,\\
\sum_{\vec k} {1 \over z_{\vec k}} \chi_{R_1}(C(\vec k))) 
\chi_{R_2}(C(\vec k)))= \delta_{R_1 R_2} ~,~~~~~~~~~~~~~~~~~~~~~~~~~~~~~~~~~~
\end{eqnarray}
where $\chi_{R_{\alpha}}(C(\vec k^{(\alpha)}))$'s are characters of the
symmetry group $S_{\ell_{\alpha}}$ with $\ell_{\alpha}=\sum_j j k_j^{(\alpha)}$
and $C(\vec k^{(\alpha)})$ are the conjugacy classes associated
to $\vec k^{(\alpha)}$'s (denoting $k_j^{(\alpha)}$ 
cycles of length $j$), 
one can show that the eqn. (\ref {compa}) becomes
\begin{equation}
Z(\{U_{\alpha}\}, \{V_{\alpha} \})=  \sum_{\{R_{\alpha}\}}
\prod_{\alpha=1}^r {\rm Tr}_{R_{\alpha}}(U_{\alpha}) {\rm Tr}_{R_{\alpha}}
(V_{\alpha})~ \label {compb}.
\end{equation} 
Ooguri and Vafa have conjectured a specific form for the vaccum expectation
value (vev) of the topological operators (\ref {opr}) for knots\cite {ov} 
invoking the large $N$ topological string duality. This result was further
refined for links\cite{laba,labb} which is generalisable for
framed links as follows 
\begin{eqnarray} 
\langle Z(\{U_{\alpha}\}, \{V_{\alpha} \}) \rangle_A&=&  
\exp\left[\sum_{d=1}^{\infty}
\sum_{\{R_{\alpha}\}}{1 \over d} 
f_{(R_1,\ldots R_r)}(q^d, \lambda^d)\prod_{\alpha=1}^r {\rm Tr}_{
R_{\alpha}} V_{\alpha}^d \right]~, \label {fexpn}\\
f_{(R_1, R_2, \ldots R_r)} (q, \lambda)&=&\lambda^{{1\over 2}\sum_{\alpha} 
\ell_{\alpha} p_{\alpha}}\sum_{Q,s} 
{1 \over (q^{1/2}- q^{-1/2})} N_{(R_1, \ldots R_r),Q,s}
q^s \lambda^Q~ \label {fexpo}
%\\
%~&=&\sum_{R_1', \ldots R_r'} M_{R_1, \ldots R_r; R_1',\ldots R_r'}
%\hat f_{( R_1',\ldots R_r')}(q,\lambda)
\end{eqnarray}
where the suffix $A$ on the vev implies that the expectation
value is obtained by integrating the $U(N)$ gauge fields $A$'s on
$S^3$, and $\ell_{\alpha}$ is the total number 
of boxes in the Young Tableau of the representation $R_{\alpha}$. 
Further, for framed links $N_{(R_1, \ldots R_r),Q,s}$ are integers 
only if expectation value of the $U(N)$ Wilson loop operators
in $S^3$ are defined as\cite {mari}
\begin{eqnarray}
\langle \prod_{\alpha=1}^r {\rm Tr}_{R_{\alpha}}(U_{\alpha}) \rangle&=&
(-1)^{\sum_{\alpha=1}^r \ell_{\alpha} p_{\alpha}} 
V_{R_1, \ldots R_r}^{\{U(N)\}}[L,S^3]~ \label {incorp}\\
~&=&\lambda^{\sum_{\alpha} {\ell_{\alpha} p_{\alpha} \over 2}} 
\tilde V_{R_1, \ldots R_r}^{\{U(N)\}}[L,S^3](q,\lambda)~. 
\end{eqnarray}
This is justified since the holonomy $V_{\alpha}$ on the
Lagrangian three-cycle ${\cal C}$, under change of framing 
becomes $\tilde V_{\alpha}= (-1)^{p_{\alpha}} V_{\alpha}$\cite{mari}, which is 
equivalent to 
\begin{equation}
{\rm Tr}_{R_{\alpha}} \tilde V_{\alpha} = (-1)^{\ell_{\alpha}p_{\alpha}}
{\rm Tr}_{R_{\alpha}} V_{\alpha}~. 
\end{equation}
If we also integrate the $\tilde A$ fields in the Chern-Simons 
field theory on the Lagrangian three-cycle then vev of the 
topological operator (\ref {compb}) is
\begin{equation}
\langle Z(\{U_{\alpha}\}, \{V_{\alpha} \}) \rangle_{A, \tilde A}=  
\sum_{R_1,\ldots R_r} \langle \prod_{\alpha=1}^r {\rm Tr}_{R_{\alpha}}
V_{\alpha} \rangle~ 
\langle \prod_{\alpha=1}^r {\rm Tr}_{R_{\alpha}}
U_{\alpha} \rangle~. \label {mostgen}
\end{equation}
where we need to determine the expectation value of
the Wilson loop operators on the Lagrangian three-cycle ${\cal C}$
with Betti number $b_1=r$  which are non-compact. 
For these non-compact Lagrangian three-cycles, 
it appears to be possible to deform knots and links into 
unknot and disjoint collection of unknots respectively. Therefore, it
is convincing to assume 
\begin{equation}
\langle {\rm Tr}_{R_{\alpha}} \tilde V_{\alpha} \rangle = {\rm dim}_q 
R_{\alpha} \equiv {S_{0 R_{\alpha}} \over S_{00}}~. \label {incorpa}
\end{equation}
Even though the assumption looks logical, finding a proof
still remains a challenging question.
Substituting eqns.(\ref {incorp}), (\ref {incorpa}) in (\ref {mostgen})
and comparing with eqns.(\ref {pinde}), (\ref {usef}), we get the relation
\begin{equation}
\alpha^{\sigma[L]} {Z_0[M] \over (S_{00})^{r+1}} = \langle Z(\{U_{\alpha}\}, 
\{V_{\alpha} \}) \rangle_{A, \tilde A}~.  \label {surg}
\end{equation}
Also, from eqn.(\ref {fexpn}), we get
\begin{equation}
{\rm ln} \left(\alpha^{\sigma[L]} {Z_0[M] \over (S_{00})^{r+1}} \right) = 
\sum_{d=1}^{\infty} \sum_{R_1, \ldots R_r}{1 \over d} 
f_{(R_1,\ldots R_r)}(q^d, \lambda^d)\prod_{\alpha=1}^r \langle {\rm Tr}_{
R_{\alpha}} V_{\alpha}^d \rangle~. \label {fina}
\end{equation}
In the limit when $N \rightarrow \infty$ and $g_s \rightarrow 0$,
the Chern-Simons partition function $Z[M]$ can be 
approximated to $Z_0[M]$ (\ref {pinde}) 
and the above equation relates the Chern-Simons free energy  
to the expectation value of the topological
link operator (\ref {opr}) in topological closed string theory.
It is appropriate to mention that the information of other 
three-manifolds obtained from surgery 
on framed links in $S^3$ in Chern-Simons theory
is captured by integrating both the $A$ and $\tilde A$ gauge fields in
the topological string theory. It will be interesting to 
see whether the Chern-Simons free energy $({\rm ln Z_0[M]})$ 
can be shown to be equal to the closed string partition function in 
these cases. We will see later that the Chern-Simons 
free energy in fact resembles the A-model partition function.

Before proceeding to the large $N$ expansion of the free energy, 
we briefly recapitulate some salient features of the A-model topological 
string partition function on Calabi-Yau manifolds. 

\subsection{The A-model Topological String Partition Function}

The A-model topological partition function on a Calabi-Yau manifold $X$
with some number of Kahler parameters denoted by $t_i$'s is defined as 
\begin{eqnarray}
\hat F(X)&=& \sum_g g_s^{2g-2} \hat {{\cal F}}_g(\{t_i\})~,\\
\hat {{\cal F}}_g(\{t_i\})&=&\sum_{\{\beta_i \}\in H_2(X, \bf Z)} 
N_{\{\beta_i\}}^g e^{-\vec \beta . \vec t}
\end{eqnarray}
where $\hat {{\cal F}}_g(\{t_i\})$ are the $A$-model topological string amplitudes
at genus $g$ and $N_{\{\beta_i\}}^g$ are the closed string Gromov-Witten invariants
associated to genus $g$ curves in the homology class $\vec {\beta}$. 
In Ref.\cite {gv3}, a strong structure result has been derived from M-theory
for the topological string partition function 
\begin{equation}
\hat F(X) = {\sum_{\{m_i\},r \geq 0,d >0}} ~~{1 \over d} ~n_{r,\{m_i\}}
(2 \sinh {d g_s \over 2})^{2r-2} \exp[-d(\sum m_i t_i)] \label {cparti}
\end{equation}
where $n_{r, \{m_i\}}$ are integers usually referred to
as Gopakumar-Vafa invariants. Clearly, 
the genus $g$ Gromov-Witten invariants $N_{\{\beta_i\}}^g$ 
will involve the set of Gopakumar-Vafa invariants 
$n_{r\leq g,\{m_i\} \leq \{\beta_i\}}$.

We would like to derive a large $N$ closed string 
expansion for the Chern-Simons free energy ${\rm ln} Z_0[M]$ 
on any three manifold $M$ (\ref {fina}) and show that it has 
the structure (\ref {cparti}).  It is not a priori clear 
whether the Chern-Simons free energy on any $M$ will have a closed 
string interpretation. Interestingly, using the duality connection 
between Chern-Simons theory on $S^3$ and topological strings and 
some properties of group theory, we will 
show that the free energy (\ref {fina}) does have the 
form (\ref {cparti}) for a subset of knots and links. 

We can write the RHS of the eqn. (\ref {fina}) as follows 
\begin{eqnarray}
\sum_{R_1, \ldots R_r} 
f_{(R_1,\ldots R_r)}(q^d, \lambda^d)\prod_{\alpha=1}^r 
\langle {\rm Tr}_{R_{\alpha}} V_{\alpha}^d \rangle= ~~~~~~~~~~~~~~~~~~~~~~~~~~~~~~~~~~~~~~~~~~~~~~~~~\nonumber \\ 
\sum_{\vec k^{(1)}, \ldots 
\vec k^{(r)}} f_{\vec k^{(1)}, \ldots \vec k^{(r)}}(q^d, \lambda^d)
\prod_{\alpha=1}^r {1 \over z_{\vec k^{(\alpha)}}} 
\langle {\bf \gamma}_{\vec k^{(\alpha)}}
(V_{\alpha}^d)~ \rangle,~~~~~~~~~~~~~~~~~~~~~~
\end{eqnarray}
where $f_{\vec k^{(1)}, \ldots \vec k^{(r)}}$ is the character
transform of $f_{R_1, \ldots R_r}$ whose form has been  
derived in \cite{laba}, namely,
\begin{eqnarray}
f_{\vec k^{(1)}, \ldots \vec k^{(r)}}(q,\lambda)&=&
\left( {\prod_j (q^{j \over 2} - q^{-{j \over 2}})
^{\sum_{\alpha=1}^r k_j^{(\alpha)}} 
\over (q^{1\over 2} - q^{-{1 \over 2}})^2}\right)
(\lambda^{1 \over 2})^{\left[\sum_{\alpha}p_{\alpha}
\left(\sum_j j k_j^{(\alpha)} \right) \right]} 
 \times \nonumber\\
~&~&\times \sum_Q \sum_{g \geq 0} n_{(\vec k^{(1)}, \ldots \vec k^{(r)}),g,Q} 
~(q^{-{1\over 2}} - q^{1 \over 2})^{2g} \lambda^Q~,
\end{eqnarray}
where 
\begin{equation}
n_{(\vec k^{(1)}, \ldots \vec k^{(r)}),g,Q} = \sum_{R_1, \ldots R_r}
\prod_{\alpha=1}^r \chi_{R_{\alpha}}(C(\vec k^{(\alpha)})) \hat N_{(R_1,
\ldots R_r),g,Q}~~. \label {hatn}
\end{equation}
$\hat N_{(R_1, \ldots R_r),g,Q}$ are integers which compute the
net number of BPS domain walls of charge $Q$ and spin $g$ transforming
in the representation $R_{\alpha}$ of $U(M)$ in the topological
string theory.
As $V_{\alpha}$'s correspond
to disjoint unknots with appropriate sign corresponding
to the framing numbers, we can write
\begin{equation}
\langle \prod_{\alpha=1}^r 
{\bf \gamma}_{\vec k^{(\alpha)}}(V_{\alpha}^d) \rangle = (-1)^{d \left[
\sum_{\alpha}p_{\alpha}
\left(\sum_j j k_j^{(\alpha)}\right) \right]} 
\left( {\prod_j (\lambda^{dj \over 2} - \lambda^{-{dj \over 2}})
\over (q^{dj\over 2} - q^{-{dj \over 2}})}\right)^{\sum_{\alpha=1}^r 
k_j^{(\alpha)}} ~.
\end{equation}
Incorporating the above results in eqn.(\ref {fina}), we get
\begin{eqnarray}
{\rm ln}\left(\alpha^{\sigma[L]} {Z_0[M] \over (S_{00})^{r+1}} \right)=
\sum_{d=1}^{\infty} \sum_{g,Q} {1 \over d} 
\left(2 \sinh {d g_s \over 2}\right)^{2g-2} 
\{ \lambda^{dQ} \times  
~~~~~~~~~~~~~~~~~~~ &~&  \label {impo}\\
\sum_{\vec k^{(1)}, \ldots \vec k^{(r)}} 
n_{(\vec k^{(1)}, \ldots \vec k^{(r)}),g,Q} 
\prod_{\alpha=1}^r \left(
{1 \over z_{\vec k^{(\alpha)}}} (- \lambda^{1 \over 2})^{d p_{\alpha}\left(
\sum_j j k_j^{(\alpha)}\right)}
\prod_j (\lambda^{-{dj \over 2}}- \lambda^{dj \over 2})^{k_j^{(\alpha)}}
\right) \} ~&~&~~~~~~~~  \nonumber
\end{eqnarray}
The RHS of the above equation has the structure of the free energy 
for a closed string (\ref {cparti}) provided we can prove that the expression 
within parenthesis satisfies
\begin{eqnarray}
\{ \lambda^{dQ}
\sum_{\vec k^{(1)}, \ldots \vec k^{(r)}} 
n_{(\vec k^{(1)}, \ldots \vec k^{(r)}),g,Q} 
\prod_{\alpha=1}^r \left(
{1 \over z_{\vec k^{(\alpha)}}}
(- \lambda^{1 \over 2})^{d p_{\alpha}\left(
\sum_j j k_j^{(\alpha)}\right)}
\prod_j (\lambda^{-{dj \over 2}}- \lambda^{dj \over 2})^{k_j^{(\alpha)}}
\right) \}
~&~&  \label {gopav}\\
= \sum_{\{m_i\}} n_{g,\{m_i\}} e^{- d\sum_i m_it_i} ~&~&  \nonumber
\end{eqnarray} 
Identifying, $\lambda=\exp(t)$, we see that 
the above relation will be true for Calabi-Yau spaces 
with one Kahler parameter after performing appropriate analytic 
continuation of the variable $\lambda \rightarrow \lambda^{-1}$
\cite{ov}. We should be able to extract the integer invariants
(the Gopakumar-Vafa invariants) from the
open string invariants $n_{(\vec k^{(1)}, \ldots \vec k^{(r)}),g,Q}$. 
Using the results obtained in Ref. \cite{laba}, one can show that
\begin{eqnarray}
(-\lambda^{1 \over 2})^{d p_{\alpha} 
\left(\sum_j j k_j^{(\alpha)}\right) }\prod_j 
(\lambda^{-{dj \over 2}}- \lambda^{dj \over 2})^{k_j^{(\alpha)}}
&=& 
%(-\lambda^{1 \over 2})^{\ell_{\alpha}p_{\alpha}d} 
(\lambda^{-{d \over 2}}- \lambda^{d \over 2}) \label {unkno}\\
~&~&\sum_{R_{\alpha}}
(-\lambda^{1 \over 2})^{\ell_{\alpha}p_{\alpha}d} 
\chi_{R_\alpha} (C(\vec k^{(\alpha)})) S_{R_{\alpha}}(\lambda^d)~, \nonumber
\end{eqnarray}
where $S_{R_{\alpha}}(\lambda)$ is zero if $R_{\alpha}$ is not a
hook representation and if $R_{\alpha}$ is a hook representation with
$\ell_{\alpha}$ boxes with $\ell_{\alpha}-s_{\alpha}$ boxes in the
first row (and we denote the representation as 
$R_{\ell_{\alpha}, s_{\alpha}}$) then we have
\begin{equation}
S_{R_{\alpha}}(\lambda^d)= 
%(-1)^{p_{\alpha} \ell_{\alpha}d} 
(-1)^{s_{\alpha}} \lambda^{d (-{\ell_{\alpha}-1 
\over 2}+ s_{\alpha})} 
%\lambda^{\ell_{\alpha} p_{\alpha} d \over 2}
\end{equation}
Using the properties(\ref {hatn}),(\ref {unkno}) in eqn.(\ref {impo})
we get
\begin{eqnarray}
{\rm ln}\left(\alpha^{\sigma[L]} {Z_0[M] \over (S_{00})^{r+1}} \right)=
\sum_{d=1}^{\infty} \sum_g {1 \over d} 
\left(2 \sinh {d g_s \over 2}\right)^{2g-2} \{\sum_Q
\sum_{\{\ell_{\alpha}\}}
\sum_{\{s_{\alpha}\}} \hat N_{(R_{\ell_1,s_1}, \ldots R_{\ell_r,s_r}),g,Q} 
\times \nonumber\\
(-1)^{\sum_{\alpha} s_{\alpha}}(-1)^{d \sum_{\alpha}\ell_{\alpha} p_{\alpha}} 
\lambda^{{1 \over 2}d \sum_{\alpha}\ell_{\alpha} p_{\alpha}}\left(
\lambda^{d\{Q+\sum_{\alpha}(-{\ell_{\alpha} \over 2}+ s_{\alpha})
\}}- \lambda^{d\{Q+1+\sum_{\alpha}(-{\ell_{\alpha} \over 2}+ s_{\alpha})\}}
\right)\}~  \nonumber \\
~=
\sum_{g,d,m}{1 \over d} (2\sinh {dg_s \over 2})^{2g-2} n_{g,m} e^{- dmt}~~~~~~~~~~~~~~~~~~~~
\label {resu}
\end{eqnarray}
We can obtain the Gopakumar-Vafa invariants $n_{g,m}$, and
also the Gromov-Witten invariants, 
by evaluating the $\hat N_{(R_1, \ldots R_r),g,Q}$ for various
framed knots and links. 

Before proceeding to outline the evaluation of the $\hat N$'s 
in the next subsection, we remind the reader that
in our computation of the Gopakumar-Vafa and the
Gromov-Witten invariants, we use the full 
Chern-Simons partition function in (\ref{resu}), and show that
this has a closed string interpretation. For reasons 
that have been outlined in the introduction, our results do not
constitute new dualities between Chern-Simons theory on $M$ and closed
A-type topological string theories, although we believe that these
results would be very important for gaining a full 
understanding of the same.  

Also, note that even though one can check that
$\sum_{\alpha}( -\ell_{\alpha}/2 +Q)$ is always an integer, 
the term  $[(-1) \lambda^{1/2}]^{\sum_{\alpha}
\ell_{\alpha} p_{\alpha}}$ within the parenthesis of 
(\ref{resu}) can be integral powers of $\lambda$ (for 
arbitrary $\ell_{\alpha}$'s) if and only if $p_{\alpha}$'s on 
the components knots are even. This suggests that the closed 
string expansion (\ref {gopav}) is possible only for framed knots and 
links with even numbers of framing numbers $p_{\alpha}$'s on 
all the component knots.

\subsection{Determination of the $\hat N$'s from framed link invariants}

The general formula for $f$ (\ref{fexpo}) in terms of framed link 
invariants (\ref {usef}) can be written as \cite{labb}
\begin{eqnarray}
f_{R_1,R_2, \ldots R_r}(q, \lambda)&=& \lambda^{\sum_{\alpha} \ell_{\alpha}
p_{\alpha}/2} \sum_{d,m=1}^{\infty} (-1)^{m-1}
{\mu(d) \over dm} \sum_{\{\vec k^{(\alpha j)},R_{\alpha j} \}}\times
\nonumber\\
~&~&\prod_{\alpha=1}^r \chi_{R_{\alpha}} \left( C\left( (\sum_{j=1}^m 
\vec k^{(\alpha j)} )_d \right) \right) \prod_{j=1}^m 
{\vert C(\vec k^{(\alpha j)})\vert \over \ell_{\alpha j}!} \times
\nonumber\\
~&~& \chi_{R_{\alpha j}}(C(\vec k^{(\alpha j)})) ~\tilde V_{R_{1j}, R_{2j},
\ldots R_{rj}}^{\{U(N)\}}[L,S^3](q^d, \lambda^d) \label {findd}
\end{eqnarray}
where $\mu(d)$ is the Moebius function defined as follows: 
if $d$ has a prime decomposition ($\{p_i\}$), $d= \prod_{i=1}^a
p_i^{m_i}$, then $\mu(d)=0$ if any of the $m_i$ is greater
than one. If all $m_i=1$, then $\mu(d)=(-1)^a$.
The second sum in the above equation runs over all vectors 
$\vec k^{(\alpha j)}$, with
$\alpha=1 , \ldots r$ and $j=1, \ldots m$, such that 
$\sum_{\alpha=1}^r \vert \vec k^{(\alpha j)} \vert > 0$ for
any $j$ and over representations $R_{\alpha j}$. 
Further $\vec k_d$ is defined as follows: $(\vec k_d)_{di}= k_i$ and has
zero entries for the other components. Therefore, if
$\vec k= (k_1, k_2, \ldots)$, then
\begin{equation}
\vec k_d=(0,\ldots, 0, k_1, 0, \ldots ,0,k_2,0,\ldots),
\end{equation}
where $k_1$ is in the $d$-the entry, $k_2$ in the $2d$-th entry, and so
on. Hence, one can directly evaluate $f$ 
from $U(N)$ framed link invariants (\ref {usef})
and verify the conjecture (\ref{fexpo}).
Using the following equations
\begin{eqnarray}
M_{R_1, \ldots R_r;R_1',\ldots R_r'} &=& \sum_{R_1'', \ldots
R_r''} \prod_{\alpha=1}^r C_{R_{\alpha} R_{\alpha}' R_{\alpha}''} 
S_{R_{\alpha}''}(q)~, \label {mrepn}\\
\hat f_{( R_1',\ldots R_r')}(q,\lambda)&=&(q^{-1/2}-q^{1/2})^{r-2}
\sum_{g \geq 0,Q} \hat N_{(R_1', \ldots R_r'),g,Q} (q^{-1/2}-q^{1/2})^{2g}
\lambda^Q~,\nonumber
\end{eqnarray}
we can write eqn.(\ref {findd}) as
\begin{equation}
f_{R_1,R_2, \ldots R_r}(q, \lambda)= \lambda^{{1\over 2}\sum_{\alpha} \ell_{\alpha}
p_{\alpha}} \sum_{R_1', \ldots R_r'}  
M_{R_1, \ldots R_r;R_1',\ldots R_r'} \hat f_{( R_1',\ldots R_r')}(q,\lambda)~.
\end{equation}
In eqn.(\ref{mrepn}), $R_{\alpha},R_{\alpha}', R_{\alpha}''$ are 
representations of the symmetric group $S_{\ell_{\alpha}}$ which can be
labelled by a Young-Tableau with a total of $\ell_{\alpha}$ boxes 
and $C_{RR'R''}$ are the Clebsch-Gordan coefficients of the symmetric
group. In the next section, we will evaluate 
$\hat N$ for few framed knots, links and present the results of our 
computation of the Gopakumar-Vafa invariants and Gromov-Witten 
invariants.

\section{Examples and explicit Results}

Our aim in this section is to compute the Gopakumar-Vafa invariants and 
Gromov-Witten invariants corresponding to the Chern-Simons free energy on 
three manifolds obtained from surgery of the respective framed links in $S^3$. 

\subsection{Knots in standard framing}

In standard framing, there is no distinction between $U(N)$ and
$SU(N)$ knot invariants.
Therefore for this class of knots with zero framing number, 
the Chern-Simons partition function will be 
\begin{equation}
Z[M]=Z_0[M]~,
\end{equation}
and the signature of the linking matrix $\sigma[L]=0$.
Substituting in eqn.(\ref {resu}), we get
\begin{eqnarray}
{\rm ln} Z[M] - 2 {\rm ln} Z[S^3]&=&
\sum_{d=1}^{\infty} \sum_g {1 \over d} 
\left(2 \sinh {d g_s \over 2}\right)^{2g-2} 
\times \label {resua}\\
~&~&\{\sum_Q
\sum_{\ell,s} \hat N_{(R_{\ell,s}),g,Q} 
(-1)^s \left( \lambda^{d(Q-{\ell \over 2}+ s)}
- \lambda^{d(Q+1-{\ell \over 2}+ s)}
\right)\}~ \nonumber\\ 
~&=&
\sum_{g,d,m}{1 \over d} (2 \sinh {dg_s \over 2})^{2g-2} n_{g,m} e^{- dmt}~~~~~~~~~~~~~~~~~~~~~~~~~ \nonumber
\end{eqnarray}
We will now compute the integer coefficients $n_{g,m}$ for few examples.

\subsubsection{Unknot} 

The surgery of this simplest knot
gives manifold $S^2 \times S^1$ whose Chern-Simons 
partition function $Z[S^2 \times S^1]=1$. 
The non-zero $\hat N$'s for the simplest unknot is \cite {ov}
\begin{equation}
\hat N_{\one,0,Q=\pm 1/2}= \mp 1 ~.
\end{equation}
Substituting this result in eqn.(\ref {resua}) and doing the
appropriate analytic continuation $\lambda \rightarrow 
\lambda^{-1}$, the non-zero $n_{g,m}$ is
\begin{equation}
n_{0,1}=2 ~.
\end{equation}
Hence eqn.(\ref {resua}) reduces to
\begin{equation}
-2 {\rm ln} Z[S^3] = \sum_{d} {1 \over d (2 \sinh {d g_s \over 2})^2} 2e^{-dt}~.
\end{equation}
Clearly, the large $N$ expansion of Chern-Simons free-energy on $S^3$ 
gives Gopakumar-Vafa invariant ($n_{g,m}[S^3]$) 
\begin{equation}
n_{0,1}[S^3]=-1~.
\end{equation}
Using the integer invariants, we can evaluate the Gromov-Witten
invariants 
\begin{equation}
N^0_{m>0}[S^3]={-1 \over m^3} ~, 
N^1_{m>0}[S^3]={1 \over 12m}~, 
N^2_{m>0}[S^3]={-m \over 240~}.
\end{equation}
These invariants in the closed topological string partition
function imply that the target Calabi-Yau space is a resolved 
conifold. 

\subsubsection{Torus knots of type (2, 2m+1)}  

The knots obtained as a closure of two-strand braid with 
$2m+1$ crossings are the type $(2,2m+1)$ torus knots.
The surgery of these torus knots in $S^3$ will give
Seifert homology spheres 
$X({2 \over -1}, {2m+1 \over m+1}, {-(2m+1) \over 1}).$
It will be interesting to determine the Gopakumar-Vafa
integer invariants and the closed Gromov-Witten invariants
corresponding to large $N$ expansion of Chern-Simons
free-energy on such Seifert manifolds.

(i) The $\hat N_{R,g,Q}$ corresponding to the torus knot (2,5)  
for represenations upto three boxes are tabulated below 
\begin{center}
\begin{tabular}{l|rrr}  \hline
\em Q & \em g=0 & \em g=1 & \em g=2\\ \hline
3/2 & 3 & 4 & 1\\
5/2 & -5 & -5 & -1\\
7/2 & 2 & 1 & 0\\ \hline
\end{tabular}
%\vskip \baselineskip

$\hat{N}_{\one,g,Q}$ for the torus knot $(2,5)$
\end{center}
\begin{center}
\begin{tabular}{l|rrrrrrrr}  \hline
\em Q & \em g=0 & \em g=1 & \em g=2 & \em g=3 & 
\em g=4 & \em g=5 & \em g=6 & \em g=7\\ \hline
3 & 20 & 60 & 69 & 38 & 10 & 1 & 0 & 0\\
4 & -80 & -260 & -336 & -221 & -78 & -14 & -1 & 0\\
5 & 120 & 400 & 534 & 366 & 136 & 26 & 2 & 0\\
6 & -80 & -260 & -336 & -221 & -78 & -14 & -1 & 0\\
7 & 20 & 60 & 69 & 38 & 10 & 1 & 0 & 0\\ \hline
\end{tabular}
%\vskip \baselineskip

$\hat{N}_{\twohor,g,Q}$ for the torus knot $(2,5)$
\end{center}
\begin{center}
\begin{tabular}{l|rrrrrrrr}  \hline
\em Q & \em g=0 & \em g=1 & \em g=2 & \em g=3 & \em g=4 & 
\em g=5 & \em g=6 & \em g=7\\ \hline
3 & 30 & 115 & 176 & 137 & 57 & 12 & 1 & 0\\
4 & -120 & -490 & -819 & -724 & -365 & -105 & -16 & -1\\
5 & 180 & 750 & 1286 & 1174 & 616 & 186 & 30 & 2\\
6 & -120 & -490 & -819 & -724 & -365 & -105 & -16 & -1\\
7 & 30 & 115 & 176 & 137 & 57 & 12 & 1 & 0 \\ \hline
\end{tabular}
%\vskip \baselineskip

$\hat{N}_{\twover,g,Q}$ for the torus knot $(2,5)$
\end{center}
\begin{center}
\begin{tabular}{l|rrrrrrr}  \hline
\em g & \em Q=9/2 & \em Q=11/2 & \em Q=13/2 & \em Q=15/2 & \em Q=17/2 & \em Q=19/2 & \em Q=21/2 \\ \hline
0 & 232 & -1652 & 4820 & -7400 & 6320 & -2852 &532 \\
1 & 1436 & -11626 & 37290 & -61400 & 55140 & -25726 & 4886 \\
2 & 4046 & -38060 & 135824 & -241510 & 228824 & -110390 & 21266 \\
3 & 6781 & -75590 & 303114 & -584700 & 584729 & -290550 & 56216 \\
4 & 7384 & -100086 & 456013 & -958591 & 1011218 & -514589 & 98651 \\
5 & 5384 & -92128 & 483836 & -1115009 & 1240265 & -642511 & 120163 \\
6 & 2636 & -60064 & 370471 & -943863 & 1107524 & -580839 & 104135 \\
7 & 851 & -27853 & 206727 & -589169 & 730275 & -385792 & 64961 \\
8 & 173 & -9107 & 83995 & -272258 & 357280 & -189269 & 29186 \\
9 & 20 & -2048 & 24548 & -92689 & 129164 & -68333 & 9338 \\
10 & 1 & -301 & 5020 & -22898 & 34006 & -17899 & 2071 \\
11 & 0 & -26 & 681 &-3984 & 6331 & -3304 & 302 \\
12 & 0 & -1 & 55 & -462 & 789 & -407 & 26 \\
13 & 0 & 0 & 2 & -32 & 59 & -30 & 1 \\
14 & 0 & 0 & 0 & -1 & 2 & -1 & 0 \\
15 & 0 & 0 & 0 & 0 & 0 & 0 & 0 \\
16 & 0 & 0 & 0 & 0 & 0 & 0 & 0 \\
17 & 0 & 0 & 0 & 0 & 0 & 0 & 0 \\ \hline
\end{tabular}
%\vskip \baselineskip

$\hat{N}_{\threehor,g,Q}$ for the torus knot $(2,5)$
\end{center}
\begin{center}
\begin{tabular}{l|rrrrrrr}  \hline
\em g & \em Q=9/2 & \em Q=11/2 & \em Q=13/2 & \em Q=15/2 & \em Q=17/2 & \em Q=19/2 & \em Q=21/2 \\ \hline
0 & 778 & -5483 & 15755 & -23750 & 19880 & -8783 & 1603 \\
1 & 5929 & -46514 & 145060 & -232875 & 204460 & -93539 & 17479 \\
2 & 20986 & -186222 & 636631 & -1095250 & 1012006 & -479592 & 91441 \\
3 & 44960 & -458386 & 1732046 & -3205225 & 3116781 & -1523901 & 293725 \\ 
4 & 64066 & -764419 & 3219215 & -6427475 & 6569210 & -3295156 & 634559 \\
5 & 63300& -905137 & 4290087 & -9275345 & 9951953 & -5092334 & 967476 \\
%\hline
%\end{tabular}
%
%$\hat{N}_{\mixed,g,Q}$ for the torus knot $(2,5)$
%\end{center}
%\begin{center}
%\begin{tabular}{l|rrrrrrr}  \hline
%\em g & \em Q=9/2 & \em Q=11/2 & \em Q=13/2 & \em Q=15/2 & \em Q=17/2 & \em Q=19/2 & \em Q=21/2 \\ \hline
6 & 44151 & -780483 & 4213699 & -9914620 & 11161326 & -5796079 &1072006 \\
7 & 21814 & -496526 & 3099128 & -7990833 & 9440203 & -4952673 & 878887 \\
8 & 7564 & -233985 & 1719912 & -4905169 & 6086843 & -3213286 & 538121 \\
9 & 1795 & -81304 &720501 & -2301847 & 3004880 & -1590473 & 246448 \\
10 & 277 & -20526 &226187 & -823741 & 1133523 & -599603 & 83883 \\
11 & 25 & -3656 & 52319 & -222684 & 323786 & -170666 & 20876 \\
12 & 1 & -435 & 8643 & -44635 & 68760 & -36018 & 3684 \\
13 & 0 & -31 & 964 & -6421 & 10510 & -5458 & 436 \\ 
14 & 0 & -1 & 65 & -626 & 1092 & -561 & 31 \\
15 & 0 & 0 & 2 & -37 & 69 & -35 & 1 \\
16 & 0 & 0 & 0 & -1 & 2 & -1 & 0 \\
17 & 0 & 0 & 0 & 0 & 0 & 0 & 0 \\ \hline
\end{tabular}
%\vskip \baselineskip 

$\hat{N}_{\mixed,g,Q}$ for the torus knot $(2,5)$.
\end{center}
\begin{center}
\begin{tabular}{l|rrrrrrr}  \hline
\em g & \em Q=9/2 & \em Q=11/2 & \em Q=13/2 & \em Q=15/2 & \em Q=17/2 & \em Q=19/2 & \em Q=21/2 \\ \hline
0 & 612 & -4282 & 12170 & -18100 & 14920 & -6482 & 1162 \\
1 & 5506 & -42271 & 129215 & -203600 & 175690 & -79121 & 14581 \\
2 & 23192 & -198031 & 657139 & -1103915 & 1000889 & -467586 & 88312 \\
3 & 59384 & -572273 & 2078169 & -3736000 & 3559209 & -1718033 & 329544 \\
4 & 101720 & -1124627 & 4504027 & -8688844 & 8685457 & -4307675 & 829942 \\
5 & 121996 & -1578123 & 7029420 & -14594746 & 15285709 & -7746221 & 1481965 \\
6 & 104816 & -1625710 & 8134469 & -18248211 & 20007137 & -10307863 & 1935362 \\
7 & 65114 & -1249175 & 7106311 & -17316684 & 19871498 & -10362997 & 1885933 \\
8 & 29205 & -721405 & 4735067 & -12623787 & 15169790 & -7976692 & 1387822 \\
9 & 9339 & -313241 & 2415866 & -7114329 & 8962560 & -4735426 & 775231 \\ 
10 & 2071 & -101529 & 941668 & -3102690 & 4104397 & -2172177 & 328260 \\
11 & 302 & -24156 & 277825 & -1042441 & 1450978 & -766965 & 104457 \\
\hline
\end{tabular}

$\hat{N}_{\threever,g Q}$ for the torus knot $(2,5)$
\end{center}
\begin{center}
\begin{tabular}{l|rrrrrrr}  \hline
\em g & \em Q=9/2 & \em Q=11/2 & \em Q=13/2 & \em Q=15/2 & \em Q=17/2 & \em Q=19/2 & \em Q=21/2 \\ \hline
12 & 26 & -4090 & 60907 & -266857 & 391757 & -206277 & 24534 \\
13 & 1 & -466 & 9605 & -51024 & 79211 & -41446 & 4119 \\
14 & 0 & -32 & 1029 & -7046 & 11600 & -6018 & 467 \\
15 & 0 & -1 & 67 & -663 & 1161 & -596 & 32 \\
16 & 0 & 0 & 2 & -38 & 71 & -36 & 1 \\
17 & 0 & 0 & 0 & -1 & 2 & -1 & 0\\ \hline
\end{tabular}
%\vskip \baselineskip

$\hat{N}_{\threever,g Q}$ for the torus knot $(2,5)$ contd.
\end{center}
The surgery of the torus knot $(2,5)$ gives the
Seifert manifold $M_1 \equiv X({2 \over -1}, {5 \over 3}, {-5 \over 1})$.
Comparing powers of $\lambda^{-1}$ (after analytic continuation) in
eqn.(\ref {resua}) we can obtain
Gopakumar-Vafa integer invariants ($n_{g,m}[M_1]$), corresponding
to large $N$ expansion of the Chern-Simons free energy on 
$M_1$, in terms of $\hat N$'s as follows 
\begin{eqnarray}
n_{g,1}[M_1]&=&\hat N_{\one,g,3/2}-2\delta_{g,0},~ \nonumber\\
n_{g,2}[M_1]&=&\hat N_{\one,g,5/2}-\hat N_{\one,g,3/2}+\hat N_{\twohor,g,3}~, \\
n_{g,3}[M_1]&=&\hat N_{\one,g,7/2}-\hat N_{\one,g,5/2}- \hat N_{\twohor,g,3}
+\hat N_{\twohor,g,4}-\hat N_{\twover,g,3} + \hat N_{\threehor,g,9/2}~. 
\nonumber
\end{eqnarray}
\begin{center}
\begin{tabular}{|l|rrrrrrrrrrrr|}  \hline
\em m & \em g=0 & \em g=1 & \em g=2 & \em g=3 & \em g=4 & \em g=5 & 
\em g=6& \em g=7&\em g=8&\em g=9
&\em g=10&\em g $\geq$ 11 \\ \hline
1 & 1 & 4 & 1 & 0 & 0 & 0 & 0&0&0&0&0&0 \\ 
2&12&51&67&38&10&1&0&0&0&0&0&0\\
3&109&1407&3466&6385&7239&5357&2634&851&173&20&1&0\\
\hline
\end{tabular}

$n_{g,m}[M_1]$
\end{center}
From these invariants, we can extract Gromov-Witten invariants
which are rational numbers. A few of them are given below 
\begin{center}
\begin{tabular}{|l|rrr|}\hline
\em m&\em g=0&\em g=1&\em g =2\\\hline
1&1&${47 \over 12}$&${241 \over 240}$\\
2&${97 \over 8}$&${1247 \over 24}$&${8287 \over 120}$\\ \hline
\end{tabular}

$N^g_m[M_1]$
\end{center}
%\begin{eqnarray}
%N^0_1[M_1]=1, ~&~&~N^0_2[M_1]= {97 \over 8}, \\
%N^1_1[M_1]={49 \over 24},~&~&~N^1_2[M_1]={1247 \over 24},\\
%N^2_1[M_1]={321 \over 320},~&~&N^2_2[M_1]={11038 \over 160}~.
%\end{eqnarray}
(ii) The $\hat N_{R,g,Q}$ computation of the torus knot $(2,7)$ 
will be
\begin{center}
\begin{tabular}{l|rrrr} \hline
\em Q & \em g=0 & \em g=1 & \em g=2 & \em g=3 \\ \hline
5/2 & -4 & -10 & -6 & -1 \\
7/2 & 7 & 14 & 7 & 1 \\
9/2 & -3 & -4 & -1 & 0 \\ \hline
\end{tabular}
%\vskip \baselineskip

$\hat{N}_{\one,g,Q}$ for the torus knot $(2,7)$
\end{center}
%\begin{center}
\begin{tabular}{l||rrrrr||} \hline
\em g & \em Q=5 & \em Q=6 & \em Q=7 & \em Q=8 & \em Q=9 \\ \hline
0 & -84 & 336 & -504 & 336 & -84 \\
1 & -574 & 2380 & -3612 & 2380 & -574 \\
2 & -1652 & 7182 & -11060 & 7182 & -1652 \\
3 & -2623 & 12144 & -19042 & 12144 & -2623 \\
4 & -2529 & 12739 & -20420 & 12739 & -2529 \\
5 & -1536 & 8673 & -14274 & 8673 & -1536 \\
6 & -589 & 3892 & -6606 & 3892 & -589 \\
7 & -138 & 1141 & -2006 & 1141 & -138 \\
8 & -18 & 210 & -384 & 210 & -18 \\
9 & -1 & 22 & -42 & 22 & -1 \\
10 & 0 & 1 & -2 & 1 & 0 \\
11 & 0 & 0 & 0 & 0 & 0 \\ \hline
\end{tabular}
%%\vskip \baselineskip
%$\hat{N}_{\twohor,g,Q}$ for the torus knot $(2,7)$
%\end{center}
\hspace{.1in}
%\begin{center}
\begin{tabular}{||rrrrr||} \hline
 \em Q=5 & \em Q=6 & \em Q=7 & \em Q=8 & \em Q=9 \\ \hline
  -112 & 448 & -672 & 448 & -112 \\
  -896 & 3696 & -5600 & 3696 & -896 \\
  -3052 & 13104 & -20104 & 13104 & -3052 \\
  -5812 & 26300 & -40976 & 26300 & -5812 \\
  -6844 & 33188 & -52688 & 33188 & -6844 \\
  -5212 & 27692 & -44960 & 27692 & -5212 \\
  -2607 & 15640 & -26066 & 15640 & -2607 \\
 -849 & 6003 & -10308 & 6003 & -849 \\
 -173 & 1541 & -2736 & 1541 & -173 \\
 -20 & 253 & -466 & 253 & -20 \\
 -1 & 24 & -46 & 24 & -1 \\
 0 & 1 & -2 & 1 & 0 \\ \hline
\end{tabular}

$\hat{N}_{\twohor,g,Q}$ for the torus knot $(2,7)$ \hspace{1in}
$\hat{N}_{\twover,g,Q}$ for the torus knot $(2,7)$

%\begin{tabular}{l|rrrrr} \hline
%\em g & \em Q=5 & \em Q=6 & \em Q=7 & \em Q=8 & \em Q=9 \\ \hline
%0 & -112 & 448 & -672 & 448 & -112 \\
%1 & -896 & 3696 & -5600 & 3696 & -896 \\
%2 & -3052 & 13104 & -20104 & 13104 & -3052 \\
%3 & -5812 & 26300 & -40976 & 26300 & -5812 \\
%4 & -6844 & 33188 & -52688 & 33188 & -6844 \\
%5 & -5212 & 27692 & -44960 & 27692 & -5212 \\
%6 & -2607 & 15640 & -26066 & 15640 & -2607 \\
%7 & -849 & 6003 & -10308 & 6003 & -849 \\
%8 & -173 & 1541 & -2736 & 1541 & -173 \\
%9 & -20 & 253 & -466 & 253 & -20 \\
%10 & -1 & 24 & -46 & 24 & -1 \\
%11 & 0 & 1 & -2 & 1 & 0 \\ \hline
%\end{tabular}
%\vskip \baselineskip

%$\hat{N}_{\twohor,g,Q}$ for the torus knot $(2,7)$ \hspace{.5in}
%$\hat{N}_{\twover,g,Q}$ for the torus knot $(2,7)$
%\end{center}
\noindent
From eqn.(\ref {resua}), we can obtain Gopakumar-Vafa integer
invariants $n_{g,m}[M_2]$ corresponding to Chern-Simons
theory on Seifert manifold $M_2=X({2 \over -1}, {7 \over 4}, {-7 \over 1}).$
We present few of them in terms of $\hat N_{R,g,Q}$ 
\begin{eqnarray}
n_{g,1}[M_2]&=&-2 \delta_{g,0} \\
n_{g,2}[M_2]&=& \hat N_{\one,g,5/2} \nonumber\\
n_{g,3}[M_2]&=& \hat N_{\one,g,7/2}- \hat N_{\one,g,5/2} \nonumber\\
n_{g,4}[M_2]&=& \hat N_{\one,g,9/2}-\hat N_{\one,g,7/2}+ \hat N_{\twohor,g,5}
\nonumber\\
n_{g,5}[M_2]&=&-\hat N_{\one,g,9/2}-\hat N_{\twohor,g,5}+
\hat N_{\twohor,g,6} - \hat N_{\twover,g,5} \nonumber
\end{eqnarray}
From these integer invariants, it is straightforward to obtain
Gromov-Witten rational numbers.

It appears from the computation of $\hat N_{R,g,Q}$'s 
for the two torus knots that the range of $Q$ is
${\ell}(2m-1)/2 \leq Q \leq \ell(2m+3)/2$. This range of $Q$
allows finite number of $\hat N$'s to contribute to $n_{g,m}, N^g_m$
invariants. Thus we see that the large $N$ expansion of
the Chern-Simons free energy on Seifert manifolds obtained
from surgery of torus knots of type $(2, 2m+1)$ 
can be given a closed string interpretation. The target
Calabi-Yau space $Y$ corresponding to the closed topological string theory
must have the Gopakumar-Vafa integer invariants and closed Gromov-Witten
invariants determined from Chern-Simons theory. So far, we 
considered only knots in standard framing. In the next subsection
we address three-manifolds obtained from framed knots in $S^3$.

\subsection{Framed Knots}

We have seen that the Chern-Simons partition function $Z[M]$,
corresponding to manifolds obtained from  knots with non-zero framing number,
can be approximated to $Z_0[M]$ in the limit $N \rightarrow \infty,$ 
$g_s \rightarrow 0$. We shall now determine $\ln Z_0[M] $ for
framed unknot with even framing number $2p$.

\noindent 
\underline{Unknot with framing $p$}: The surgery of $p$-framed
unknot results in Lens spaces ${\cal L}(p,1)$. 
The $\hat N$'s for the unknot with framing $p=4$ are  
\begin{center}
\begin{tabular}{l|r} \hline
\em Q & \em g=0 \\ \hline
-1/2 & 1 \\
1/2 & -1 \\ \hline
\end{tabular}
%\vskip \baselineskip

$\hat{N}_{\one,g,Q}$ for the unknot with framing $p=4$
\end{center}

\begin{center}
\begin{tabular}{l|rrrr} \hline
\em Q & \em g=0 & \em g=1 & \em g=2 & \em g=3 \\\hline
-1 & 2 & 1 & 0 & 0 \\
0 & -6 & -5 & -1 & 0 \\
1 & 4 & 4 & 1 & 0 \\ \hline
\end{tabular}
\hspace{1in}
\begin{tabular}{l|rrrr} \hline
\em Q & \em g=0 & \em g=1 & \em g=2 & \em g=3 \\\hline
-1 & 4 & 4 & 1 & 0 \\
0 & -10 & -15 & -7 & -1 \\
1 & 6 & 11 & 6 & 1 \\ \hline
\end{tabular}
%\vskip \baselineskip

$\hat{N}_{\twohor,g,Q}$ for the unknot with $p=4$ \hspace{1in} 
$\hat{N}_{\twover,g,Q}$ for the unknot with $p=4$
\end{center}

\begin{center}
\begin{tabular}{l|rrrrrrrrrrr} \hline
\em Q & \em g=0 & \em g=1 & \em g=2 & \em g=3 & \em g=4 & \em g=5 & \em g=6 & \em g=7 & \em g=8 & \em g=9 & \em g=10 \\ \hline
-3/2 & 12 & 26 & 22 & 8 & 1 & 0 & 0 & 0 & 0 & 0 & 0 \\
-1/2 & -58 & -181 & -246 & -175 & -67 & -13 & -1 & 0 & 0 & 0 & 0 \\
1/2 & 86 & 335 & 582 & 550 & 298 & 92 & 15 & 1 & 0 & 0 & 0 \\
3/2 & -40 & -180 & -358 & -383 & -232 & -79 & -14 & -1 & 0 & 0 & 0 \\ \hline
\end{tabular}

$\hat{N}_{\threehor,g,Q}$ for the unknot with framing $p=4$
\end{center}

\begin{center}
\begin{tabular}{l|rrrrrrrrrrr} \hline
\em Q & \em g=0 & \em g=1 & \em g=2 & \em g=3 & \em g=4 & \em g=5 & \em g=6 & \em g=7 & \em g=8 & \em g=9 & \em g=10 \\ \hline
-3/2 & 46 & 155 & 224 & 167 & 66 & 13 & 1 & 0 & 0 & 0 & 0 \\
-1/2 & -206 & -915 & -1836 & -2057 & -1377 & -561 & -136 & -18 & -1 & 0 & 0 \\
1/2 & 290 & 1545 & 3768 & 5226 & 4446 & 2394 & 817 & 171 & 20 & 1 & 0 \\
3/2 & -130 & -785 & -2156 & -3336 & -3135 & -1846 & -682 & -153 & -19 & -1 & 0 \\ \hline
\end{tabular}

$\hat{N}_{\mixed,g,Q}$ for the unknot with framing $p=4$
\end{center}

\begin{center}
\begin{tabular}{l|rrrrrrrrrrr} \hline
\em Q & \em g=0 & \em g=1 & \em g=2 & \em g=3 & \em g=4 & \em g=5 & \em g=6 & \em g=7 & \em g=8 & \em g=9 & \em g=10 \\ \hline
-3/2 & 40 & 180 & 358 & 383 & 232 & 79 & 14 & 1 & 0 & 0 & 0 \\
-1/2 & -170 & -965 & -2514 & -3719 & -3367 & -1925 & -696 & -154 & -19 & -1 & 0 \\
1/2 & 230 & 1535 & 4746 & 8446 & 9374 & 6748 & 3196 & 987 & 191 & 21 & 1 \\
3/2 & -100 & -750 & -2590 & -5110 & -6239 & -4902 & -2514 & -834 & -172 & -20 & -1 \\ \hline
\end{tabular}

$\hat{N}_{\threever,g,Q}$ for the unknot with framing $p=4$
\end{center}
Using these $\hat N_{R,g,Q}$, we can determine Gopakumar-Vafa 
integer invariants $n_{g,m}[{\cal L}(4,1)]$ from 
the expansion of $\ln Z_0[{\cal L}(p,1)]$. We present some
of the non-zero coefficients 
\begin{eqnarray}
n_{g,1}[{\cal L}(4,1)]&=&\hat N_{\one,g,-1/2} \delta_{g,0}-2 \delta_{g,0}~,\\
n_{g\leq 1,2}[{\cal L}(4,1)]&=&\hat N_{\one,g,1/2}-\hat N_{\one,g,-1/2}+ \hat N_{\twohor,g,-1}~,
\nonumber\\
n_{g\leq 4,3}[{\cal L}(4,1)]&=&-\hat N_{\one,g,1/2}+ \hat N_{\twohor,g,0}-
\hat N_{\twohor,g,-1}- \hat N_{\twover,g,-1}+\hat N_{\threehor,g,-3/2}~,
\nonumber\\
n_{g,4}[{\cal L}(4,1)]&=&\hat N_{\twohor,g,1}- \hat N_{\twohor,g,0}- 
\hat N_{\twover,g,0}+ \hat N_{\twover,g,-1}+ \hat N_{\threehor,g,-1/2}~,
\nonumber\\
&~&-\hat N_{\threehor,g,-3/2}- \hat N_{\mixed,g,-3/2}+ \hat N_{\fourhor,g,-2}~,
\nonumber
\end{eqnarray}
and the closed Gromov-Witten rational numbers can be
deduced from the integer invariants. 
In the appendix, we have the $\hat N_{R,g,Q}$ for 
representations upto two boxes for the unknot with arbitrary framing.
They will be useful to determine $n_{g,m}$ corresponding
to Chern-Simons theory on Lens spaces ${\cal L}(2p,1)$.
In the following subsection, we will consider framed links.

\subsection{Framed Links}

We take Hopf link $H^*(p_1,p_2)$ with linking number
$\ell k=-1$ and the framing on the two component knots
as $p_1=p_2=4$. Surgery of such a framed link in $S^3$
will give Lens space ${\cal L}(15,4)$. The 
$\hat N_{(R_1,R_2),g,Q}= \hat N_{(R_2,R_1),g,Q}$ for this example 
is tabulated below:
\begin{center}
\begin{tabular}{l|r} \hline
\em Q & \em g=0 \\ \hline
-1 & 1 \\
0 & -1 \\ \hline
\end{tabular}
%\vskip \baselineskip

$\hat{N}_{(\one,\one),g,Q}$ for the framed Hopf link $H^*(4,4)$
\end{center}

\begin{center}
\begin{tabular}{l|rrrr} \hline
\em Q & \em g=0 & \em g=1 & \em g=2 & \em g=3\\ \hline
-3/2 & 3 & 1 & 0 & 0 \\
-1/2 & -9 & -6 & -1 & 0 \\
1/2 & 6 & 5 & 1 & 0 \\ \hline
\end{tabular}
\hspace{0.5in}
\begin{tabular}{l|rrrr} \hline
\em Q & \em g=0 & \em g=1 & \em g=2 & \em g=3\\ \hline
-3/2 & 6 & 5 & 1 & 0 \\
-1/2 & -16 & -20 & -8 & -1 \\
1/2 & 10 & 15 & 7 & 1 \\ \hline
\end{tabular}
%\vskip \baselineskip

$\hat{N}_{(\twohor,\one),g,Q}$\ for $H^*(4,4)$
\hspace{.5in}$\hat{N}_{(\twover,\one),g,Q}$ for $H^*(4,4)$
\end{center}

\begin{center}
\begin{tabular}{l|rrrrrrrrrrr} \hline
\em Q & \em g=0 & \em g=1 & \em g=2 & \em g=3 & \em g=4 & \em g=5 & \em g=6 & \em g=7 & \em g=8 & \em g=9 & \em g=10 \\ \hline
-2 & 23 & 41 & 29 & 9 & 1 & 0 & 0 & 0 & 0 & 0 & 0 \\
-1 & -117 & -312 & -367 & -230 & -79 & -14 & -1 & 0 & 0 & 0 & 0 \\
0 & 180 & 606 & 920 & 771 & 376 & 106 & 16 & 1 & 0 & 0 & 0 \\
1 & -86 & -335 & -582 & -550 & -298 & -92 & -15 & -1 & 0 & 0 & 0 \\ \hline
\end{tabular}
%\vskip \baselineskip

$\hat{N}_{(\threehor,\one),g,Q}$ for the framed Hopf link $H^*(4,4)$
\end{center}

\begin{center}
\begin{tabular}{l|rrrrrrrrrrr} \hline
\em Q & \em g=0 & \em g=1 & \em g=2 & \em g=3 & \em g=4 & \em g=5 & \em g=6 & \em g=7 & \em g=8 & \em g=9 & \em g=10 \\ \hline
-2 & 94 & 271 & 338 & 221 & 78 & 14 & 1 & 0 & 0 & 0 & 0 \\
-1 & -438 & -1697 & -3001 & -3003 & -1820 & -680 & -153 & -19 & -1 & 0 & 0 \\
0 & 634 & 2971 & 6431 & 8008 & 6188 & 3060 & 969 & 190 & 21 & 1 & 0 \\
1 & -290 & -1545 & -3768 & -5226 & -4446 & -2394 & -817 & -171 & -20 & -1 & 0 \\ \hline
\end{tabular}
%\vskip \baselineskip

$\hat{N}_{(\mixed,\one),g,Q}$ for the framed Hopf link $H^*(4,4)$
\end{center}
\begin{center}
\begin{tabular}{l|rrrrrrrrrrr} \hline
\em Q & \em g=0 & \em g=1 & \em g=2 & \em g=3 & \em g=4 & \em g=5 & \em g=6 & \em g=7 & \em g=8 & \em g=9 & \em g=10 \\ \hline
-2 & 86 & 335 & 582 & 550 & 298 & 92 & 15 & 1 & 0 & 0 & 0 \\
-1 & -376 & -1880 & -4350 & -5776 & -4744 & -2486 & -832 & -172 & -20 & -1 & 0 \\
0 & 520 & 3080 & 8514 & 13672 & 13820 & 9142 & 4013 & 1158 & 211 & 22 & 1 \\
1 & -230 & -1535 & -4746 & -8446 & -9374 & -6748 & -3196 & -987 & -191 & -21 & -1 \\ \hline
\end{tabular}
%\vskip \baselineskip

$\hat{N}_{(\threever,\one),g,Q}$ for the framed Hopf link $H^*(4,4)$
\end{center}

\begin{center}
\begin{tabular}{l|rrrrrrr} \hline
\em Q & \em g=0 & \em g=1 & \em g=2 & \em g=3 & \em g=4 & \em g=5 & \em g=6 \\ \hline
-2 & 15 & 17 & 7 & 1& 0 & 0 & 0 \\
-1 & -60 & -83 & -45 & -11 & -1 & 0 & 0 \\
0 & 81 & 126 & 75 & 20 & 2 & 0 & 0 \\
1 & -36 & -60 & -37 & -10 & -1 & 0 & 0 \\ \hline
\end{tabular}
%\vskip \baselineskip

$\hat{N}_{(\twohor,\twohor),g,Q}$ for the framed Hopf link $H^*(4,4)$
\end{center}

\begin{center}
\begin{tabular}{l|rrrrrrr} \hline
\em Q & \em g=0 & \em g=1 & \em g=2 & \em g=3 & \em g=4 & \em g=5 & \em g=6 \\ \hline
-2 & 27 & 45 & 30 & 9 & 1 & 0 & 0 \\
-1 & -105 & -206 & -165 & -66 & -13 & -1 & 0 \\
0 & 138 & 301 & 262 & 113 & 24 & 2 & 0 \\
1 & -60 & -140 & -127 & -56 & -12 & -1 & 0 \\ \hline
\end{tabular}
%\vskip \baselineskip

$\hat{N}_{(\twohor,\twover),g,Q}$ for the framed Hopf link $H^*(4,4)$
\end{center}

\begin{center}
\begin{tabular}{l|rrrrrrr} \hline
\em Q & \em g=0 & \em g=1 & \em g=2 & \em g=3 & \em g=4 & \em g=5 & \em g=6 \\ \hline
-2 & 48 & 106 & 99 & 47 & 11 & 1 & 0 \\
-1 & -184 & -466 & -501 & -287 & -91 & -15 & -1 \\
0 & 236 & 660 & 767 & 470 & 159 & 28 & 2 \\
1 & -100 & -300 & -365 & -230 & -79 & -14 & -1 \\ \hline
\end{tabular}
%\vskip \baselineskip

$\hat{N}_{(\twover,\twover),g,Q}$ for the framed Hopf link $H^*(4,4)$
\end{center}
When one of the representations is trivial, the $\hat N_{({\bf .}, R),g,Q}$'s 
will be equal to the invariants $\hat N_{R,g,Q}$'s computed for unknot with 
framing $p=4$. 

The Gopakumar-Vafa invariants
from the expansion of $\ln Z_0[{\cal L}(15,4)]$ are 
\begin{eqnarray}
n_{g,1}[{\cal L}(15,4)]&=&2 n_{g,1}[{\cal L}(4,1)]+ \delta_{g,0}~,\\
n_{g,2}[{\cal L}(15,4)] &=&\delta_{g,0} \hat N_{(\one,\one),g,-1}
+ 2 n_{g,2}[{\cal L}(4,1)]~,
\nonumber\\
n_{g \leq 1,3}[{\cal L}(15,4)] &=&\hat N_{(\one,\one),g,0} -
\hat N_{(\one,\one),g,-1}~+2\hat N_{(\twohor,\one),g,-3/2}
+ 2 n_{g,3}[{\cal L}(4,1)]~,
\nonumber\\
n_{g \leq 4,4}[{\cal L}(15,4)]&=&-\hat N_{(\one,\one),g,0} 
- 2\hat N_{(\twohor,\one),g,-3/2} + 2\hat N_{(\twohor,\one),g,-1/2}
\nonumber\\ 
&~& -2\hat N_{(\twover,\one),g,-3/2}+2\hat N_{(\threehor,\one),g,-2}
+\hat N_{(\twohor,\twohor),g,-2}
\nonumber\\
&~&+ 2 n_{g,4}[{\cal L}(4,1)]~,
\nonumber 
\end{eqnarray}
These examples suggest that 
the Chern-Simons partition function on various manifolds can
be given a A-model closed string theory interpretation.
From the prediction of Gopakumar-Vafa integer invariants
it should be possible to determine the nature of the Calabi-Yau
background.

\section{Summary and Conclusions}

We have studied $U(N)$ Chern-Simons gauge theory
and framed link invariants in $S^3$, for specific
choice of $U(1)$ representations placed on the component
knots of the link. From these $U(N)$ framed link invariants in
$S^3$, we have constructed three-manifold invariants which are the
same as $SU(N)$ three-manifold invariants.
These invariants are proportional to the 
Chern-Simons partition function $Z[M]$ on the corresponding 
three-manifolds.
 
In this paper, we have used the results of the topological string
duality conjecture relating Chern-Simons theory
on $S^3$ to closed $A$ model topological string theory on the 
resolved conifold, to obtain large $N$ expansions of the
Chern-Simons free-energy ($\ln Z[M]$) on some non-trivial manifolds.
The closed string theory expansion resembles the $A$-model
topological string theory on a Calabi-Yau space with one Kahler
parameter. We have computed the Gopakumar-Vafa integer invariants
and the Gromov Witten invariants associated with Chern-Simons partition
function on three-manifolds like Seifert manifolds, Lens spaces etc.

We need to understand some subtle issues about the Chern-Simons 
partition function on any three-manifold $M$. As we have pointed out
in the introduction, the classical solutions of 
the Chern-Simons action are the flat connections on $M$, 
and in the weak coupling (large $k$) limit, the partition function, 
which may be a sum or integral over the space of flat connections,
can be written as
\begin{equation}
Z[M] = \sum_{c} Z_c[M] \equiv \int_{\mu_c} Z_c[M]~,
\end{equation}
where $Z_c[M]$ is obtained from perturbative expansion
around a stationary point $A=A_c$. The large $N$ expansion
proposed by `t Hooft requires ${\rm ln} Z_c[M]$ to have
a closed string interpretation whereas we have
shown in this paper that ${\rm ln}Z[M]$ has a closed string expansion
for many manifolds.

Note that for the case of the three-sphere $S^3$, there is only one 
stationary point, which is the trivial connection. Therefore 
$Z[S^3]$ is also equal to the perturbative expansion around the 
trivial connection and hence the closed string interpretation is
expected from `t Hooft's formulation. 

In the context of Lens spaces ${\cal L}(p,1)$, the space of flat connections 
is a set of points. In ref. \cite{akmv}, $Z[M]$ has been rewritten 
as a sum over all flat connections, enabling the extraction
of the perturbative partition function around a non-trivial
flat connection ($Z_c[M]$) for ${\cal L}(p,1)$. 
It has been shown from the matrix model 
approach that $Z_c[M]$ can be given a closed string
theoretic interpretation. The results establish the  
duality between Chern-Simons theory 
on Lens spaces ${\cal L}(p,1)$ and closed string theory on a 
$A_{p-1}$ singularity fibred over $P^1$.
The Gopakumar Vafa integer invariants that we have computed 
for Lens spaces ${\cal L}(2p,1)$ correspond to 
${\rm ln}\left( \sum_c Z_c[M]\right)$. There must be some relation 
between these integer invariants and the corresponding invariants 
on $A_{2p-1}$ singularity fibred over $P^1$ at some
special values of the Kahler parameters. We hope to 
decipher such interesting relations in future.

It appears to be a difficult task  to  rewrite $Z[M]$ 
as a sum or integral over flat connections to determine
$Z_c[M]$ for other three-manifolds like Seifert-manifolds.
The challenge lies in determining $Z_c[M]$ and the closed
string expansion to precisely state new duality 
conjectures between Chern-Simons theory on $M$ and
closed string theory. We leave the study of 
these aspects for a future publication.

\newpage
\noindent
{\bf Acknowledgements}  

\noindent
PR would like to thank M. Marino and C. Vafa 
for discussions during the initial stages of the project, and is
grateful to N. Habegger for comments. She would like to thank the 
Adbus Salam ICTP for providing local hospitality and an excellent 
academic atmosphere during her visit under the ICTP Junior Associateship 
scheme. We would like to thank S. Govindarajan, T. R Govindarajan, 
M. Marino, K. Ray and G. Thompson for discussions and clarifications. 
PB would like to thank CSIR for the grant.

\newpage
\appendix{\noindent\bf\Large{Appendix}}\\

\noindent
\section{{\large{$\hat N_{R,g,Q}$ for the unknot with arbitrary framing $p$}}}

\noindent
For an unknot with arbitrary framing $p$, the $\hat N_{R,g,Q}$ for 
fundamental representation is
\begin{equation}
\hat N_{\one, 0,Q=\pm 1/2}=\mp (-1)^p~, \hat N_{\one, g\neq 0,Q}=0~.
\end{equation}
For representations involving two boxes in the Young-Tableau,
$\hat N$'s for arbitrary $g$ can be written as follows 
\begin{eqnarray}
\hat N_{\twohor,p-s,-1}&=& \{\frac{\theta(s-3)}{(s-3)!}
(2p-2s+2)(2p-2s+3) \ldots (2p-s-2) \nonumber\\
&~& + \frac{\theta(s-5)}{(s-5)!}
(2p-2s+2)(2p-2s+3) \ldots (2p-s-4) + \ldots \} \nonumber\\
&~& + x\theta(s-2)~,\\
\hat N_{\twohor,p-s,0}&=&-\delta_{s,2}- \frac{\theta(s-2)}{(s-2)!}
(2p-2s+3)(2p-2s+4) \ldots (2p-s)\\
\hat N_{\twohor,p-s,1}&=&\{ \frac{\theta(s-2)}{(s-2)!}
(2p-2s+2)(2p-2s+3) \ldots (2p-s-1) \nonumber\\
&~& + \frac{\theta(s-4)}{(s-4)!}
(2p-2s+2)(2p-2s+3) \ldots (2p-s-3) + \ldots \} \nonumber\\
&~& + y\theta(s-1)~,\\
\hat N_{\twover,p-s,-1}&=& \{ \frac{\theta(s-2)}{(s-2)!}
(2p-2s+2)(2p-2s+3) \ldots (2p-s-1) \nonumber\\
&~& + \frac{\theta(s-4)}{(s-4)!}
(2p-2s+2)(2p-2s+3) \ldots (2p-s-3) + \ldots \} \nonumber\\
&~& + y\theta(s-1)~,\\
\hat N_{\twover,p-s,0}&=& - \delta_{s,1}- \frac{\theta(s-1)}{(s-1)!}
(2p-2s+3)(2p-2s+4) \ldots (2p-s+1)~,\\
\hat N_{\twover,p-s,1}&=& \{ \frac{\theta(s-1)}{(s-1)!}
(2p-2s+2)(2p-2s+3) \ldots (2p-s)\nonumber\\
&~& + \frac{\theta(s-3)}{(s-3)!}
(2p-2s+2)(2p-2s+3) \ldots (2p-s-2) + \ldots \}\nonumber\\
&~& + x\theta(s)~,
\end{eqnarray}
with $x=1$ \& $y=0$ for odd $s$,  ~$x=0$ \& $y=1$ for even $s$
and $\theta(u)$ is the usual theta function which is equal to one
if $u > 0$ and zero if $u \leq 0$. 
The negative framing ($-p$) BPS integers are related to the positive 
framing integers \cite {laba} 
\begin{equation}
\hat N^{(p)}_{(R_1, \ldots R_r),g,Q}=(-1)^{\sum_{\alpha} \ell_{\alpha}-1}
\hat N^{(-p)}_{(R^t_1, \ldots R^t_r),g,-Q}~,
\end{equation}
where $R^t_{\alpha}$'s are obtained by transposing the rows and
columns in the Young-Tableau representations. These 
integers are consistent with Marino-Vafa results
for $g=0,1,2$ \cite {mari}.

\end{document}